\documentclass[prb,aps,twocolumn,showpacs]{revtex4-1}
\usepackage{dcolumn}
\usepackage{bm}

\usepackage[colorlinks,hyperindex, urlcolor=blue, linkcolor=blue,citecolor=black, linkbordercolor={.7 .8 .8}]{hyperref}
\usepackage{graphicx}
\usepackage{amsfonts}
\usepackage{amsmath}
\usepackage{amssymb}
\usepackage{amsbsy}
\usepackage{nicefrac}

\newcommand{\rxx}{$R_{xx}$}
\newcommand{\rxy}{$R_{xy}$}
\newcommand{\sampa}{{\it A}}
\newcommand{\sampb}{{\it B}}
\newcommand{\dvv}{$\Delta v/v$}

\begin{document}

\title {Charge metastability and hysteresis in the quantum Hall regime}

\author{J. Pollanen}
\altaffiliation[Present address: ]{Department of Physics and Astronomy, Michigan State University, East Lansing, MI 48824-2320, USA}
\affiliation{Institute of Quantum Information and Matter, Department of Physics, California Institute of Technology, Pasadena, CA 91125}
\author{J.P. Eisenstein}
\affiliation{Institute of Quantum Information and Matter, Department of Physics, California Institute of Technology, Pasadena, CA 91125}
\author{L.N. Pfeiffer}
\affiliation{Department of Electrical Engineering, Princeton University, Princeton, NJ 08544}
\author{K.W. West}
\affiliation{Department of Electrical Engineering, Princeton University, Princeton, NJ 08544}

\date{\today}

\begin{abstract}
We report simultaneous quasi-dc magnetotransport and high frequency surface acoustic wave measurements on bilayer two-dimensional electron systems in GaAs.  Near strong integer quantized Hall states a strong magnetic field sweep hysteresis in the velocity of the acoustic waves is observed at low temperatures. This hysteresis indicates the presence of a metastable state with anomalously high conductivity in the interior of the sample.  This non-equilibrium state is not revealed by conventional low frequency transport measurements which are dominated by dissipationless transport at the edge of the 2D system. We find that a field-cooling technique allows the equilibrium charge configuration within the interior of the sample to be established. A simple model for this behavior is discussed.  
\end{abstract}


\maketitle

\section{Introduction}
Hysteresis can be found in numerous physical systems and situations ranging from ferromagnetic materials to the elastic response of rubber bands\cite{bro96, bert98}. When a system is subjected to a time-varying force, hysteresis results from the formation, and relaxation, of long-lived metastable configurations and is often connected with phase transitions between competing ground states\cite{bro96}. For example, in the case of a ferromagnetic material subjected to a changing magnetic field, the the dynamics of spin polarized domains leads to a history dependent magnetization.

Magnetic field sweep hysteresis has also been reported in a wide variety of experiments on single and bilayer two-dimensional electron systems (2DES) in semiconductor heterostructures in both the integer and fractional quantum Hall regimes. These hysteretic effects have been associated with a number of physical phenomena, including non-equilibrium charge distributions \cite{zhu00, pan05, siddiki06}, long-lived eddy currents \cite{jpe85, huels04, piorolad06, ell06} within the interior of the 2DES, first order phase transitions involving the electron spin \cite{kron98, cho98, kron99, eom00, poortere00, smet01} or pseudospin \cite{piazza99, jung01, tut03, misra08} degrees of freedom, and metastable orientations of the electron nematic phases at high Landau level occupancy \cite{cooper04}.

In the majority of these experiments the behavior of the collective electron state was inferred from measurements of magnetoresistance, which was found to depend strongly on the sweep rate and direction of the magnetic field. For the incompressible states of the integer and fractional quantum Hall effects the magnetoresistance is dominated by dissipationless edge channels at the perimeter of the sample and is only weakly influenced by the charge state in the interior of the 2DES. Therefore, in several experiments a complementary probe, such as a magnetometer\cite{jpe85, piorolad06, ell06} or a single electron transistor\cite{huels04} positioned within the interior of the sample, has been used to study the bulk electronic states.

In this paper we report on a new observation of magnetic field sweep dependent hysteresis in the integer quantum Hall regime. This hysteresis is manifest in the velocity of high frequency surface acoustic waves (SAWs) propagating across the bulk of a 2DES in GaAs double quantum well (DQW) samples. In these experiments the SAWs directly probe changes in the the bulk conductivity $\sigma_{xx}$ of the 2DES and allow us to observe a metastable charge state created by a controlled magnetic field sweep. This metastable bulk state is not detected by simultaneous low frequency conventional magnetotransport measurements. 

\section{SAW-2DES interaction}
Surface acoustic waves can be a sensitive probe of the conductivity of the 2DES in GaAs/AlGaAs heterostructures and have been successfully used to study the integer\cite{wixforth86, wixforth89, drichko15} and fractional\cite{willet90} quantum Hall regimes, as well as the 2D metal-to-insulator transition at zero magnetic field \cite{tracy06}. Additionally, enhanced absorption of acoustic energy can occur when the wavelength of the SAW becomes comparable to some other length scale within the 2DES. This is the case for collective states such as the quantum Hall nematics\cite{kuk11} or the Wigner solid\cite{paa92} for which the areal charge density is spatially modulated, or near Landau level filling factor $\nu= 1/2$ where the wavelength of high frequency SAWs becomes commensurate with the cyclotron orbits of composite fermions\cite{willet93}.

The interaction between a SAW and a 2DES is mediated by the electric field that propagates in tandem with the elastic wave. Charges in the 2DES attempt to dynamically screen this electric field, which shifts the SAW velocity \cite{wixforth86, simon96,velocitynote}
\begin{equation}
\frac{\Delta v}{v} = \frac{K^2}{2}\frac{1}{1+(\sigma_{xx}(q, \omega)/\sigma_{m})^2}.
\end{equation}
In Eq. 1 $\sigma_{xx}(q,\omega)$ is the longitudinal component of the magnetoconductivity tensor at wavevector $q$ and frequency $\omega$, while $K^2$ and $\sigma_m$ are material and device-dependent parameters that encode the strength of the coupling between the SAW and the 2DES and the ability of the 2DES to screen the SAW electric field.  For our samples, we find \cite{sawparam} $K^2\approx 4.6 \times 10^{-5}$ and estimate $\sigma_m \approx 8\times 10^{-7}$ $\Omega^{-1}$.  For the present work, both $q$ and $\omega$ are sufficiently small that $\sigma_{xx}(q,\omega)$ can be replaced by the ordinary Drude conductivity $\sigma_{xx}$.

As the conductivity passes from $\sigma_{xx} \gg \sigma_m$ to $\sigma_{xx} \ll \sigma_m$, the SAW velocity shift ranges from zero to its maximum value of $K^2/2$, with the greatest sensitivity to small changes in the conductivity around $\sigma_{xx}\sim \sigma_m$.  The relative smallness of $\sigma_m$ makes SAW measurements an especially effective tool in the quantum Hall effect regime where $\sigma_{xx}$ tends to extremely low values.

We emphasize that SAW measurements probe the conductivity $\sigma_{xx}$ throughout the $bulk$ of the 2DES.  This contrasts with ordinary transport measurements, especially in the quantum Hall effect regime where transport along the edge of the 2DES dominates.  For example, in the presence of a strong quantized Hall (QH) state, the insulating bulk of the 2DES is surrounded by dissipationless chiral edge channels. No voltage drop exists between contacts positioned along the sample edge (away from the current source and drain).  Moreover, this absence of an edge voltage drop will persist even if there are isolated regions in the interior of the 2DES that, owing to density fluctuations, are not robustly quantized.  The conductivity of these regions will be finite and, if their total area is not negligible, the SAW velocity shift will register their presence. 

\section{Experimental}
\subsection{Samples}
The samples used in this work are GaAs/AlGaAs double quantum wells grown via molecular beam epitaxy (MBE). Two independently grown bilayer samples of this type, referred to as \sampa\ and \sampb, have been examined. Each sample contains two 18 nm wide GaAs quantum wells separated by a 10 nm wide barrier of nominally\cite{Albarrier} Al$_{0.9}$Ga$_{0.1}$As. These quantum wells are embedded in thick cladding layers of the alloy Al$_{0.32}$Ga$_{0.68}$As. Silicon $\delta$-doping sheets are located in these cladding layers\cite{setbacks} and donate electrons that populate the lowest subband of each of the quantum wells at low temperature. As grown, sample \sampa\ has a total electron density $n_{T} = 0.85 \times 10^{11}$ cm$^{-2}$, which is nominally equally divided between the two quantum wells, and a low temperature, zero field, mobility of $0.6 \times 10^{6}$ cm$^{2}$/Vs. Sample \sampb\ has a somewhat higher total density of $1.25 \times 10^{11}$  cm$^{-2}$ but a lower mobility of $0.3 \times 10^{6}$ cm$^{2}$/Vs.

\subsection{Transport measurements}
Both DQW samples consist of $\sim5 \times 5$ mm square chips cleaved from their parent MBE wafers. Standard optical lithography was used to confine the active area of the bilayer to square mesas 2 mm on a side. Six evaporated AuNiGe Ohmic contacts were positioned along two opposing sides of the mesa and were diffused into the epilayers to contact both quantum wells in parallel. The samples were chemically thinned from the backside to a thickness of $\sim100$ $\mu$m and global aluminum top and back gates controlled the electron densities within each of the individual quantum wells. These gates can be used to balance the total electron density between the two quantum wells and, when desired, fully deplete one or the other 2D layer to create a single layer system.

The samples were thermally anchored to the mixing chamber of a dilution refrigerator and a magnetic field perpendicular to the quantum wells was supplied by a superconducting solenoid. The longitudinal and Hall resistances \rxx\ and \rxy\ were simultaneously measured using standard low frequency (13 Hz) ac lock-in techniques. For all the measurements the excitation current was $\leq 10$ nA to avoid Joule heating of the electrons. To estimate $\sigma_{xx}$ we use a classical model \cite{sigmaxx} of the current distribution in our samples to extract the resistivity $\rho_{xx}$ from the measured resistances \rxx\ and \rxy.  Inverting the resistivity tensor then provides $\sigma_{xx}$.

\subsection{Surface Acoustic Wave Measurements}
In addition to the low frequency measurements of $\sigma_{xx}$ we also simultaneously measure the velocity shift of SAWs propagating across the surface of the sample.  A matched set of two aluminum interdigitated transducers were fabricated photolithographically onto either side of the mesa\cite{sawohmic}. One of these transducers functions as a transmitter while the second detects the SAW after it interacts with the 2DES. A phase-locked loop employing a standard homodyne receiver was used to measure the velocity shift of the SAWs after they propagated across the mesa\cite{wixforth89, tracy07}. The fundamental frequency of the transducers is $\approx149$ MHz\cite{funlambda}, however the presence of the highly conducting aluminum top gate over the mesa strongly screens the interaction between the surface acoustic wave and the 2DES at this relatively low frequency \cite{wixforth89, tracy06}. To overcome this effect, and improve the signal to noise ratio of our measurements, we used the 5$^{th}$ transducer harmonic at $\approx 747$ MHz for all of our experiments. SAW excitation levels were kept sufficiently low to ensure linear response of the received signal and to avoid electron heating, which we could infer from our simultaneous low frequency magnetotransport measurements. In particular, we found that even when the base temperature of the mixing chamber was $\approx 15$ mK the minimum temperature of the 2DES was limited to 50 mK, even for our lowest SAW excitation power.

\section{Results and Discussion}
\subsection{Hysteresis in the quantum Hall regime}
Figure 1(a) shows the longitudinal conductivity $\sigma_{xx}$ of sample \sampa\ as a function of perpendicular magnetic field, $B$, deduced from the measured resistances \rxx\ and \rxy. In these measurements, done at $T = 50$ mK, both the top and back gates were grounded and the total density of the bilayer was equal to its as-grown value $n_{T} = 0.85 \times 10^{11}$ cm$^{-2}$, which was nominally equally divided between the two quantum wells. This equal division was deduced both from the absence of beating in the low field Shubnikov-de Haas oscillations of \rxx\ and from the lack of any splitting in fractional quantum Hall minima in \rxx\ at high magnetic field.

Deep minima in $\sigma_{xx}$ are observed when the 2DES exhibits the integer or fractional quantum Hall effects. In Fig. 1 (a) these QH states are labeled by the total Landau level filling factor $\nu_{T}=\nu_{1}+\nu_{2}=n_{T}h/eB$, where $\nu_{1}$ and $\nu_{2}$ are the individual layer filling factors. For example, the minimum at $B=1.79$ T corresponds to the integer QH state at total filling factor $\nu_{T}=2$. Also labeled in Fig. 1(a) are the integer state at $\nu_{T} = 4$ and the fractional QH states at $\nu_{T} = 8/3$ and $4/3$.
\begin{figure}[b]
\begin{center}
\includegraphics[width=1 \columnwidth]{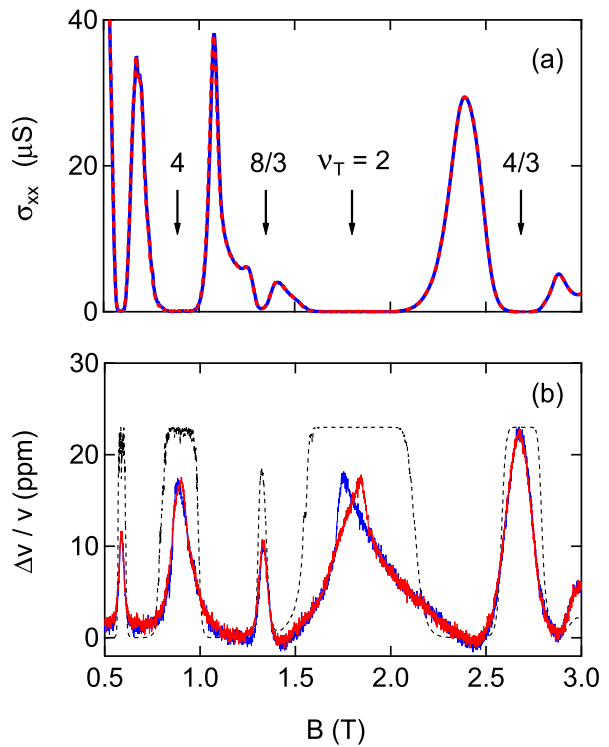}
\end{center}
\caption{(Color online) Low frequency $\sigma_{xx}$ and SAW velocity shift signatures of the quantum Hall effect in sample \sampa\ at $T=50$ mK, comparing magnetic field sweeps up (red) and down (blue). (a) Conductivity $\sigma_{xx}$ deduced from the measured resistances \rxx\ and \rxy. The downward arrows indicate the positions of various integer and fractional quantum Hall states. (b) Simultaneously measured SAW velocity shift versus magnetic field. Pronounced field-sweep hysteresis is apparent at $\nu_{T}=2$.  A weaker hysteresis is also observed at $\nu_{T}=4$. The dotted black curve shows the velocity shift calculated using Eq. 1 and the conductivity data in panel (a).}
\end{figure}

In Fig. 1(b) the simultaneously measured SAW velocity shift is plotted over the same magnetic field range as the conductivity data in Fig. 1(a). As qualitatively expected\cite{wixforth89,simon96} from Eq. 1, an increase in the SAW velocity occurs when the bilayer enters into an incompressible quantum Hall state and the conductivity $\sigma_{xx}$ falls sharply.  Within a deep quantum Hall state the velocity shift should attain its maximum value of $\Delta v/v=K^2/2$ and remain there so long as $\sigma_{xx}<<\sigma_m$.  This expectation is illustrated in Fig. 1(b) by the dotted curve which is the velocity shift calculated using Eq. 1 and the conductivity data in Fig. 1(a). Instead of plateaus, the measured velocity shift shows relatively narrow peaks at each quantum Hall state.  This discrepancy between the SAW velocity shift calculated from low frequency conventional transport measurements has, to our knowledge, been observed in all previous SAW measurements in the quantum Hall effect regime.  Since the velocity shift in the flanks of the peaks falls short of the plateau value, the implication is that the average conductivity in the bulk of the 2DES is larger than the low frequency transport data suggests.  This is consistent with the presence of conducting puddles of quasi-electrons and quasi-holes arising from the inevitable fluctuations in the 2DES density about its average value.
 
More interestingly, Fig. 1(b) reveals a strong hysteresis in the SAW velocity shift around $\nu_T=2$ depending on whether the magnetic field is swept up (red) or down (blue).  A weaker, but still observable, hysteresis is seen around $\nu_T=4$.  No hysteresis is seen at the fractional QH states at $\nu_T=4/3$ and 8/3.  In contrast, the low frequency conductivity data shown in Fig. 1(a) shows no sign of any hysteresis at $\nu_T=2$ or 4, but the very small magnitude of $\sigma_{xx}$ in these integer quantum Hall effects makes it difficult to be certain about this. For the measurements shown in Fig. 1 the magnetic field was swept at a rate of $\pm$50 mT/min; reducing the sweep rate to $\pm$5 mT/min had no effect on the hysteresis. Qualitatively similar results are observed in sample \sampb.

Figure 2 shows how the SAW velocity shift, and the hysteresis at integer filling, evolve with increasing temperature in sample \sampa.  Already at $T=100$ mK the hysteresis is significantly reduced relative to the $T=50$ mK data; by $T = 300$ mK it is absent.   Beyond the suppression of the hysteresis at $\nu_T=2$, raising the temperature has several other important effects on the SAW velocity shift.   First, the peaks in \dvv\ at $\nu_T=4/3$ and 8/3 disappear almost entirely by $T=300$ mK.  This is as expected: the small energy gaps for these fractional QH states leads to $\sigma_{xx}$ rising rapidly with temperature and once $\sigma_{xx}\gg \sigma_m$, the SAW velocity shift is essentially zero.  Second, the data in Fig. 2 show that the peaks in \dvv\ at $\nu_T=2$ and 4 narrow considerably as $T$ rises.  This too is not surprising: the conductivity of the quasiparticles present near these integer filling factors increases with $T$, thereby reducing \dvv.

\begin{figure}
\begin{center}
\includegraphics[width=1 \columnwidth]{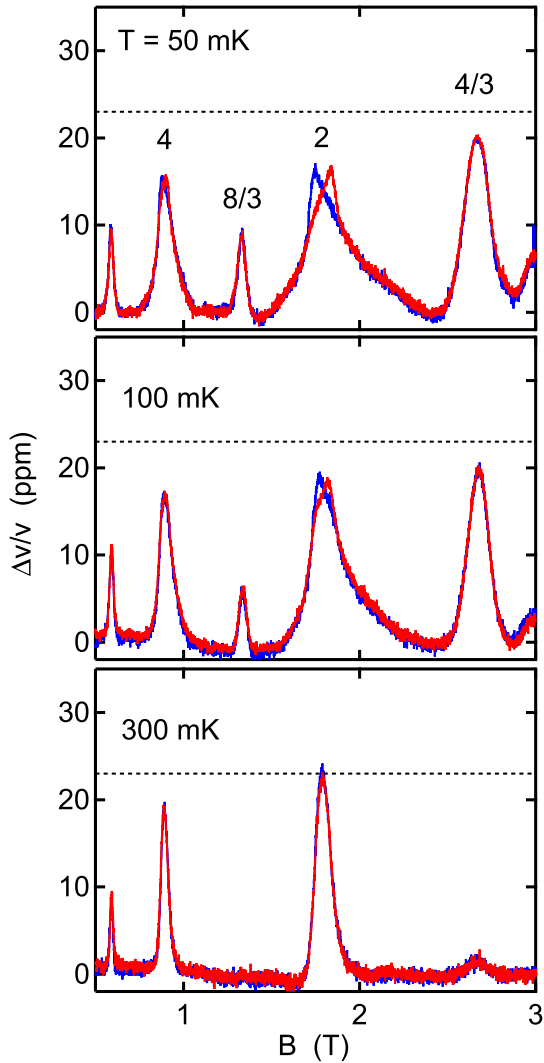}
\end{center}
\caption{(Color online) Temperature dependence of the SAW velocity shift in sample \sampa\ as a function of magnetic field with both quantum wells populated.  The hysteresis at $\nu_{T}=2$ and 4 disappears as the temperature is increased above $\approx 100$ mK. Simultaneously, the peak value of the velocity shift at these filling factors increases with increasing temperature and approaches the maximum value expected for a fully insulating 2DES in our sample (dashed horizontal line). At each temperature the magnetic field was swept up (red) and down (blue) at a rate of $\pm$50 mT/min.}
\end{figure}

What $is$ surprising in Fig. 2 is the increase of the peak heights at $\nu_T=2$ and 4 as the temperature rises.  By $T=300$ mK the $\nu_T=2$ peak has reached the maximum value expected for a fully insulating 2DES in our sample: $\Delta v/v \approx 23$ ppm.   Though not shown in the figure, the $\nu_T=2$ peak height remains at this maximum value to at least $T=500$ mK.  This observation implies that, on average across the bulk of the 2DES, the conductivity $\sigma_{xx}$ at $\nu_T=2$ and 4 is $higher$ at $T=50$ mK than it is at 300 mK and 500 mK.

We suggest that these unexpected results reflect the creation, by the magnetic field sweep, of metastable non-equilibrium states of the 2DES at the lowest temperatures.  In support of this idea, we now describe a field-cooling technique which allows the observation, in the SAW velocity shift, of the equilibrium state of the 2DES at the same low temperatures.

\subsection{Equilibrium state via field-cooling}
That elevated temperatures remove the magnetic field sweep hysteresis in the SAW velocity shift around $\nu_T=2$ and 4, and expose a smaller average conductivity at these fillings,  suggests a way to access the equilibrium state of the 2DES at low temperatures.  First, the magnetic field is swept, at an appropriately high temperature, to a chosen fixed value.  With the field held constant, the temperature is then reduced and the SAW velocity shift recorded.  This field-cooling process is then repeated at various magnetic fields of interest.  This discrete set of \dvv\ data points are then compared to field-swept low temperature velocity shift data.  Figure 3 presents this comparison, with the solid dots acquired via field-cooling from $T=500$ mK to $T=50$ mK.  The starting temperature, $T=500$ mK, was chosen since no SAW hysteresis is observed at this high temperature.

Figure 3 demonstrates that the $T=50$ mK field-swept and field-cooled \dvv\ data agree essentially perfectly at all magnetic fields where hysteresis is absent in the field-swept data.  This includes fields where the 2DES is highly conducting, $e.g.$ at $B=1.1$ T between the $\nu_T=4$ and $\nu_T=8/3$ QH states, and at the $\nu_T=4/3$ and 8/3 fractional QH states where the 2DES is relatively insulating.  In contrast, near the center of the strong $\nu_T=2$ and 4 integer QH states, where hysteresis in the field-swept \dvv\ data is seen at low temperatures, the field-cooled velocity shift is seen to be larger than the field-swept value.  Indeed, at $\nu_T=2$ the field-cooled \dvv\ value reaches $\approx 24$ ppm, within error the maximum possible value.  These data demonstrate that through field-cooling the 2DES attains a state of lower average conductivity than it does via a field sweep at the same low temperature.  We suggest that that the field-cooled low temperature state is the equilibrium one and that low temperature field sweeps to strong QH states produce a non-equilibrium configuration of the 2DES with anomalously high average conductivity.

\begin{figure}
\begin{center}
\includegraphics[width=1 \columnwidth]{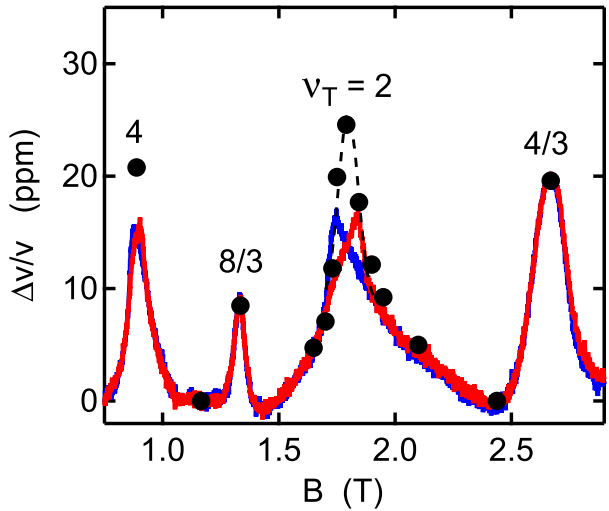}
\end{center}
\caption{(Color online) SAW velocity shift as a function of magnetic field for sample \sampa. The solid black dots were achieved by cooling the sample from $500$ mK to $50$ mK at fixed field as described in the text. These field-cooled data points can be compared with the $50$ mK field-swept data (up sweep red, down sweep blue), which exhibit strong hysteresis in the vicinity of integer filling. The dashed line is a guide to the eye.}
\end{figure}

\subsection{A simple model}
We speculate that the occurrence of the non-equilibrium charge states of the 2DES in strong QH states reported here may be qualitatively understood as follows.  If the 2DES were completely isolated and unable to exchange charge with any external reservoirs, its total density would remain fixed.  In this case the Fermi level of the 2DES will exhibit a sawtooth-like oscillation, descending through the Landau level spectrum as a magnetic field is applied.  Alternatively, if the 2DES remains in electrochemical equilibrium with a charge reservoir, then the Fermi level will remain fixed as charge is exchanged back and forth between the 2DES and the reservoir \cite{zawadzki14}.  As a function of magnetic field, the 2DES density $n$ will oscillate about its zero field value $n_0$.  On crossing a quantum Hall plateau, the density will pass from $n<n_0$ on the low magnetic field side of the plateau, to $n>n_0$ on the high field side.   The magnitude of the net charge density variation $\delta n$ depends on the capacitance between the 2DES and the reservior and the energy gap of the QH state; typically $\delta n \ll n_0$.

Since the conductivity $\sigma_{xx}$ becomes extremely small in a strong QH state, the density of the 2DES, at least in the interior of the sample, may be unable to remain in equilibrium with the reservoir as the magnetic field is being swept and become trapped at a non-equilibrium value.  For example, as the QH state is approached from the low magnetic field side, the rapidly falling conductivity might isolate the interior of the 2DES at a density less than the equilibrium value.  Conversely, if the QH state is approached from the high field side, the interior might become isolated at a density higher than equilibrium requires.  Either way, as the plateau is being crossed the deviation of the density from its equilibrium value will continue to increase until the conductivity returns to a large enough value that charge transfer with the reservoir can restore equilibrium.  Although we lack a quantitiative model for this hysteretic effect, it seems probable that the deviations of the 2DES density in the interior of the sample from its equilbrium value would, in the QH plateau region, result in an anomalously high effective conductivity.  This would explain why the observed SAW velocity shift at low temperatures fails to reach the maximum value of $\Delta v/v =K^2/2$ in the center of strong QH states.  Moreover, at higher temperatures the increased conductivity would prevent the isolation of the interior of the 2DES.  Consistent with our observations, hysteresis would disappear and, unless the temperature is too high, the SAW velocity shift would rise.

We emphasize that this charge trapping effect need not occur uniformly across the entire sample.  The inevitable inhomogeneity of the 2DES density can lead to small isolated patches of conducting quasielectrons and quasiholes entirely surrounded by strongly insulating channels of incompressible quantum Hall fluid.  Unable to exchange charge with the reservoir, these patches would be out of equilibrium and, inside strong QH states, lead to hysteresis in, and a reduced magnitude of, the net SAW velocity shift of the sample.

Finally, we note that this qualitative model does not depend upon the identification of the charge reservoir itself.  Indeed, in the literature various possible reservoirs are discussed.  For example, in their early work on this subject, Baraff and Tsui \cite{baraff81} suggested that the ionized donor impurities in a modulation-doped GaAs/AlGaAs heterostructure could play the role of the reservoir via their weak tunnel coupling with the 2DES at the heterointerface.  In the present case, we suggest instead that the In ohmic contacts and the metal gates evaporated on both surfaces of the our samples are good candidates for the required charge reservoir.   These gates are held at fixed potential relative to the ohmic contacts and, therefore, the 2DES Fermi level.   With the top surface gate about $d=0.6$ $\mu$m away from the 2DES, a crude estimate suggests that $\delta n/n_0 \sim 2\times 10^{-3}$, assuming a QH energy gap $\Delta = 10$ K.

\subsection{Single versus double layer SAW response} 
Since our samples contain a double layer 2DES, charge transfer between them, either through tunneling or via the ohmic contacts, can also occur.  In effect, each layer might play the role of charge reservoir for the other layer.  For the data discussed thus far, the densities of the two layers were nominally equal, with both the front and back gates held at ground potential.  Ignoring local density variations, the Fermi levels of the two layers move in tandem as a function of magnetic field and no chemical potential imbalances develop between the two layers as the magnetic field is varied.  That a strong SAW hysteresis is nonetheless observed, suggests that the required charge reservior is external to the double layer 2DES.  This conclusion is buttressed by further measurements of the SAW velocity shift in sample A when small density differences, $\Delta n/n \lesssim 0.08$, between the layers were intentionally imposed by appropriately biasing the gates.  No qualitative change in the velocity shift hysteresis was found.

To determine whether the presence of two 2DES layers in close proximity is essential to the hysteresis phenomena reported here, SAW velocity shift measurements were made after fully depleting the top 2DES layer using the top surface gate.  To determine the value of top gate voltage $V_{TG}$ needed for this, the two-terminal conductance of the device is measured as a function of $V_{TG}$.  As $V_{TG}$ is made increasingly negative, the two-terminal conductance remains roughly constant until the top layer nears full depletion, at which point the conductance rapidly drops to a lower plateau value.  Further increases of the (negative) top gate voltage begin to reduce the density of the lower 2DES.  We chose $V_{TG}$ to be slightly more negative than the minimum needed to deplete the upper 2DES and then applied a small positive voltage to the back gate to replenish the density of the lower 2DES to the value it had when both layers were equally populated.  Under these conditions, the SAW velocity shift is measured as a function of magnetic field; Fig. 4 shows the results for samples A and B.  

Panels (a) and (c) of Fig. 4 show the SAW velocity shift around $\nu_T=1+1=2$ in samples A and B, respectively, for the density balanced, double layer case.  These $T=50$ mK data show that a very similar hysteresis is observed in the two samples under these conditions.  In contrast, panels (b) and (d) reveal different results for the two samples when the top 2DES has been depleted.  In this situation sample A reveals only a very small SAW hysteresis around $\nu_T=1+0=1$, while sample B shows a robust effect, comparable to that seen when both layers are populated.  

The different results upon depletion of the top 2DES in samples A and B is puzzling.  These samples are structurally very similar, but differ somewhat in the their carrier concentrations and zero field mobility.  Nevertheless, the clear observation of strong SAW velocity shift hysteresis in sample B in its single layer configuration demonstrates that two layers are not required for a magnetic field sweep to induce a non-equilibrium charge state in the bulk of the 2DES at strong integer QH states.  At the same time, the lack of SAW hysteresis in sample A after the top 2DES is depleted suggests that the effect is sensitive to conditions not controlled in the present experiment.  For example, the two layers in sample A might differ significantly in the strength of their respective $\nu = 1$ QH states owing to different disorder landscapes.  A comprehensive low frequency transport study, using separate ohmic contacts to the individual 2DES layers, might reveal such differences. 

\begin{figure}
\begin{center}
\includegraphics[width=1 \columnwidth]{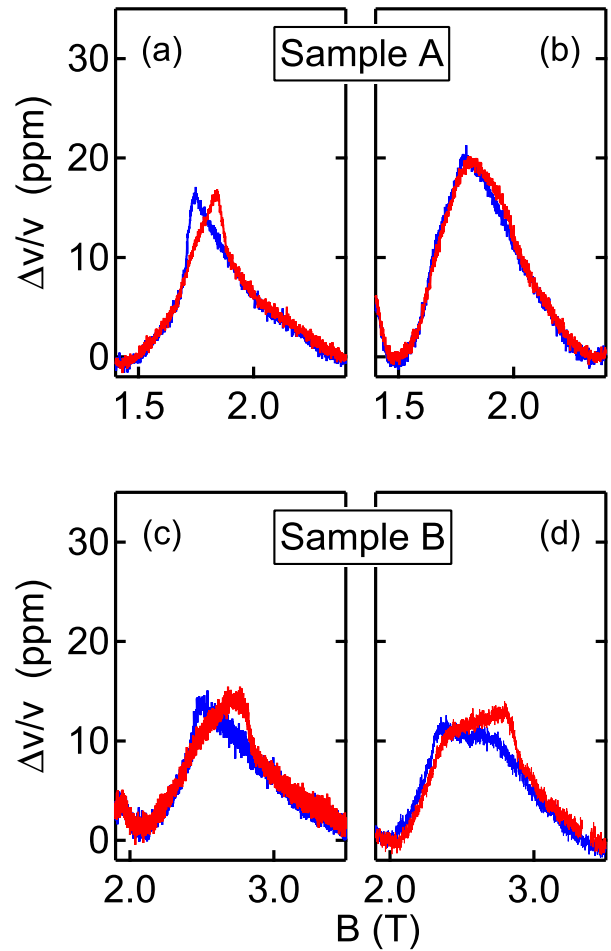}
\end{center}
\caption{(Color online) SAW velocity shift versus magnetic field at $T = 50$ mK for samples \sampa\ (top) and \sampb\ (bottom). As described in the text, for these data only the lower quantum well in each sample is populated with electrons. The density of electrons in the lower quantum well of sample \sampa\ (\sampb) is $0.43 \times 10^{11}$ cm$^{-2}$ ($0.61 \times 10^{11}$ cm$^{-2}$). These values of density were chosen so that $\nu=1$ for the lower layer occurs at approximately the same magnetic field as $\nu_{T}=2$ when both quantum wells are populated with the same density of electrons.}
\end{figure}

\section{Conclusion}
In summary, surface acoustic wave velocity shift measurements reveal an intriguing magnetic field sweep-dependent hysteresis at strong integer quantized Hall states.  This effect, which is not detectable in ordinary edge-dominated low frequency transport measurements, suggests that at low temperatures the bulk of the 2D electron system can be trapped in a very long-lived non-equilibrium state with anomalously high conductivity.   The effect disappears at elevated temperatures and this allows the non-equilibrium state to be avoided by cooling the sample at fixed magnetic field rather than adjusting the field at low temperature.  We speculate that this non-equilbrium state occurs because the extremely low conductivity of the 2DES in a strong quantum Hall state isolates the interior of the 2DES from charge reservoirs with which it is normally in equilibrium.

\begin{acknowledgements}
We thank E.H. Fradkin, B.I. Halperin, H.R. Krishnamurthy, M. Shayegan, S.H. Simon, J.C.W. Song and A. Stern for discussions. The Caltech portion of this work was supported by NSF Grant DMR-0070890, DOE grant FG02-99ER45766 and the Institute for Quantum Information and Matter, an NSF Physics Frontiers Center with support of the Gordon and Betty Moore Foundation through Grant No. GBMF1250. The work at Princeton was partially funded by the Gordon and Betty Moore Foundation through Grant GBMF2719, and by the National Science Foundation MRSEC-DMR-0819860 at the Princeton Center for Complex Materials.
\end{acknowledgements}


\begin{references}
\bibitem{sawparam}The value of $K^2$ is measured by observing the SAW velocity shift as the bilayer 2DES is fully depleted using the top gate.  The value of $\sigma_m$ is calculated from theory, taking account of the distance ($d\approx 600$ nm) of the 2DES beneath the metallized sample surface and the SAW wavelength ($\lambda\approx 3.8$ $\mu$m).
\bibitem{Albarrier} The precise aluminum concentration in the barrier between the quantum wells was 88\% in sample \sampa\ and 91\% in sample \sampb.
\bibitem{setbacks} The doping of sample \sampa\ is slightly asymmetric with the Si layers are positioned 205 nm above and 230 nm below the quantum wells. In sample \sampb\ the doping is symmetric with a set back distance of 280 nm both above and below the quantum wells.
\bibitem{sigmaxx} In our experimental configuration, current $I$ is injected and withdrawn at two adjacent corner contacts on the square mesa.  The voltage difference $V_{xx}$ between the remaining two corner contacts determines the longitudinal resistance $R_{xx}=V_{xx}/I$.  Assuming a classical model of the current distribution, the longitudinal resistivity $\rho_{xx}$ is related to the resistance via $\rho_{xx}=4.531 \times R_{xx}$.
\bibitem{sawohmic} No ohmic contacts are in the line-of-sight between the two SAW transducers.
\end{references}
\end{document}